\documentclass[aps,showpacs]{revtex4}
\begin{document}
\title{On the gravitational field induced by static electromagnetic sources}
\author{Boyko V. Ivanov}
\email{boyko@inrne.bas.bg}
\affiliation{Institute for Nuclear Research and Nuclear Energy,\\
Tzarigradsko Shausse 72, Sofia 1784, Bulgaria}

\begin{abstract}
It is argued that static electric or magnetic sources induce
Weyl-Majumdar-Papapetrou solutions for the metric of spacetime. Their
gravitational acceleration includes a term many orders of magnitude stronger
than usual perturbative terms. It gives rise to a number of effects, which
can be detected experimentally. Two electrostatic and two magnetostatic
examples of physical set-ups with simple symmetries are proposed. The
different ways in which mass sources enter and complicate the pure
electromagnetic picture are described.
\end{abstract}

\maketitle

\section{Introduction}

In classical Newton-Maxwell physics the electromagnetic (EM) fields have no
influence upon gravity, which is generated by sources of mass. In general
relativity EM fields alter the metric of spacetime and induce a
gravitational force through their energy-momentum tensor 
\begin{equation}
T_{\;\nu }^\mu =\frac 1{4\pi }\left( F^{\mu \alpha }F_{\nu \alpha }-\frac 14%
\delta _{\;\nu }^\mu F^{\alpha \beta }F_{\alpha \beta }\right) ,  \label{one}
\end{equation}
where 
\begin{equation}
F_{\mu \nu }=\partial _\mu A_\nu -\partial _\nu A_\mu  \label{two}
\end{equation}
is the electromagnetic tensor and $A_\mu $ is the four-potential. $T_{\;\nu
}^\mu $ enters the r.h.s. of the Einstein equations 
\begin{equation}
R_{\;\nu }^\mu =\kappa T_{\;\nu }^\mu .  \label{three}
\end{equation}
We have taken into account that $T_{\;\mu }^\mu =0$. The Einstein constant
is 
\begin{equation}
\kappa =\frac{8\pi G}{c^4}=2.07\times 10^{-48}s^2/cm.g,  \label{four}
\end{equation}
where $G=6.674\times 10^{-8}cm^3/g.s^2$ is the Newton constant and $%
c=2.998\times 10^{10}cm/s$ is the speed of light. We shall use the Gauss
system (CGS) of nonrelativistic units and occasionally the international
system of practical units (SI).

In addition, the Maxwell equations are coupled to gravity through the
covariant derivatives of $F_{\mu \nu }$%
\begin{equation}
F_{\quad ;\nu }^{\mu \nu }=\frac 1{\sqrt{-g}}\left( \sqrt{-g}F^{\mu \nu
}\right) _\nu =-\frac{4\pi }cJ_\mu ,\qquad J^\mu =\sigma cu^\mu .
\label{five}
\end{equation}
Here $g$ is the metric's determinant, usual derivatives are denoted by
subscripts, $J^\mu $ is the four-current, $u^\mu =dx^\mu /ds$ is the
four-velocity of the charged particles with charge density $\sigma $. We
shall study mainly electrovacuum solutions with $\sigma \neq 0$ only on some
surface, specifying the boundary conditions. The Einstein-Maxwell equations
(3,5) show how the EM-field leaves its imprint on the metric, which has to
satisfy the Rainich conditions \cite{one,two,three}.

The gravitational force acting on a test particle is represented by the
four-acceleration 
\begin{equation}
g_\mu =c^2\frac{du_\mu }{ds}=c^2\Gamma _{\alpha ,\mu \beta }u^\alpha u^\beta
=\frac{c^2}2g_{\alpha \beta ,\mu }u^\alpha u^\beta ,  \label{six}
\end{equation}
where $\Gamma _{\mu \beta }^\alpha $ are the Christoffel symbols. When the
particle is at rest, $u^0=\left( g_{00}\right) ^{-1/2}$ and 
\begin{equation}
g_\mu =\frac{c^2}2\left( \ln g_{00}\right) _\mu  \label{seven}
\end{equation}
for an arbitrary metric $g_{\alpha \beta }$.

In this paper we investigate the problem whether EM-fields can induce strong
enough acceleration, rising above the gravimeter's threshold of $%
10^{-6}cm/s^2$ or even comparable to the mean Earth acceleration $%
g_e=980.665cm/s^2$. Eqs. (3,4) show that the metric will be very near to the
flat one without any singularities and faraway from the metric of a black
hole. The question is whether the 20 orders of magnitude supplied by $c^2/2$
in Eq. (7) are enough to lift the EM-gravitational force to that of the
Newtonian gravity of very massive bodies. In fact, we should consider the
contravariant physical (tetrad) four-vector $g^{\left( \mu \right) }=\eta
^{\mu \nu }g_{\left( \nu \right) }$, where $\eta _{\mu \nu }=diag\left(
1,-1,-1,-1\right) $, but in cartesian coordinates and for such an almost
flat metric it is indistinguishable from $g_\mu $ or $g_{\left( \mu \right)
} $ except for a sign change.

It seems natural to use perturbation theory in the harmonic gauge where 
\begin{equation}
g_{\mu \nu }=\eta _{\mu \nu }+h_{\mu \nu },\qquad \Delta h_{\mu \nu
}=-2\kappa T_{\mu \nu }  \label{eight}
\end{equation}
and the Poisson equation shows that $h_{\mu \nu }$ is extremely small, while 
$g_\mu \sim c^2\kappa =1.85\times 10^{-25}CGS$. It appears that the problem
has been solved and the effect is negligible. However, there is a loophole,
which is discussed in the sequel.

In Sec. II it is shown that the gravitational acceleration in
Weyl-Majumdar-Papapetrou (WMP) fields \cite{five,eleven,twelve,thirt,fourt}
includes a term proportional to $\sqrt{\kappa }$ and linear in the fields.
In Sec. III axially-symmetric static Einstein-Maxwell fields are revisited.
The three classes of Weyl solutions are described, as well as their relation
to the general solution and situations where they become the most general
solution. It is argued that charge and current distributions determine their
harmonic master-potential and induce them as pure electromagnetic effect
upon the spacetime metric. A general formula is given for the acceleration
and the question of hidden mass sources is investigated. In Sec. IV a short
review is given of WMP fields in the general static case and in the presence
of charged dust or perfect fluid.

In Sec. V we give two electrostatic examples of charged surfaces, which
induce Weyl solutions. They involve plane, spherical and spheroidal
symmetry. The charged plane and the Reissner-Nordstr\"om (RN) solution are
discussed from a Weyl point of view. The peculiarities in the junction
conditions for Weyl fields are pointed out. Different physical issues, such
as the force arising inside a charged capacitor and repulsive gravity around
a charged sphere are explored. In Sec. VI we give two magnetostatic
examples, a current loop and an open solenoid. In Sec. VII the main results
about WMP fields are summarised. The last section contains a short
discussion.

\section{Root gravity}

Let us assume that the metric and the EM-fields do not depend on time. In
this stationary case let us further simplify the problem by setting $A_\mu
=\left( \bar \phi ,0,0,0\right) $. There is just an electric field 
\begin{equation}
E_\mu =F_{0\mu }=-\bar \phi _\mu .  \label{eleven}
\end{equation}
Obviously, $T_{\;\nu }^\mu $ from Eq. (1) contains only quadratic terms in $%
\bar \phi _\mu $. This allows to hide $\kappa $ from Eq. (3) by normalizing
the electric potential to a dimensionless quantity 
\begin{equation}
\phi =\sqrt{\frac \kappa {8\pi }}\bar \phi .  \label{twelve}
\end{equation}
The factor $8\pi $ is chosen for future convenience. We shall see that this
is a much more elegant way to get rid of the constants in the
Einstein-Maxwell equations than the choice of relativistic units $c=1$ and $%
G=1$ ($8\pi G=1$ sometimes).

Imagine now that in some exact solution $g_\mu $ is proportional to the
electric field, contrary to the quadratic dependence in Eq. (8) 
\begin{equation}
g_\mu =ac^2\phi _\mu =ac^2\sqrt{\frac \kappa {8\pi }}\bar \phi _\mu ,
\label{thirt}
\end{equation}
where $a$ is some slowly changing function of order $O\left( 1\right) $. For
example, when $a=a\left( g_{00}\right) $ then Eqs. (7,11) lead to the
functional dependence $f\equiv g_{00}=F\left( \phi \right) $. Let us further
assume that the spacetime is static. Then the above functional dependence
has the unique form \cite{eleven,twelve,thirt} 
\begin{equation}
f=1+B\phi +\phi ^2.  \label{fourt}
\end{equation}
In the axially-symmetric case this equation was found by Weyl already in
1917 \cite{fourt} and such solutions are known as Weyl fields. The potential
in Eq. (10) is very small everywhere and naturally goes to zero at infinity.
Then asymptotic flatness fixes the first term which is otherwise an
arbitrary constant. It is also fixed by the requirement to go back to
Minkowski spacetime when $\phi =0$ since we are studying the EM-effect on
gravity with no masses present. Later we will give arguments that the
typical value of $B$ is $2$, so that the linear term in Eq. (12) is really
present.

Thus in Weyl-Majumdar-Papapetrou fields we have 
\begin{equation}
g_\mu =c^2f^{-1}\left( \frac B2\sqrt{\frac \kappa {8\pi }}\bar \phi _\mu +%
\frac \kappa {8\pi }\bar \phi \bar \phi _\mu \right) .  \label{fift}
\end{equation}
The first term is of the type given in Eq. (11), the second resembles the
expression in Eq. (8). Let us note that 
\begin{equation}
c^2\sqrt{\frac \kappa {8\pi }}=\sqrt{G}=2.58\times 10^{-4},\qquad c^2\frac 
\kappa {8\pi }=\frac G{c^2}=7.37\times 10^{-27}.  \label{sixt}
\end{equation}
Due to the square root, the first coefficient is $10^{23}$ times bigger than
the second. We shall call gravitational fields, which have acceleration
terms $\sim \sqrt{\kappa }$, root gravity. The WMP fields are an example,
but there are others too. Thus general relativity has a Newtonian limit in
the case of mass sources, where $g_\mu \sim G$ and a Maxwellian limit in the
case of EM-sources, where $g_\mu \sim \sqrt{G}$. In relativistic units this
effect is not seen. When $G=c=1$ then $\sqrt{\kappa /8\pi }=\kappa /8\pi =1$
and there is no difference between root and usual gravity. When $8\pi G=c=1$
then $\kappa =1$ and $\sqrt{\kappa /8\pi }=0.2,$ while $\kappa /8\pi =0.04$.
The difference is just an order of magnitude.

Provided that $B=2$ our search for a strong gravitational acceleration
induced by EM-fields doesn't seem so doomed as in the introduction.
Electrostatic generators can create potential differences of six million
volts or 
\begin{equation}
\bar \phi _{\max }=2\times 10^4CGS.  \label{sevent}
\end{equation}
If it were applied to a capacitor with distance of $1cm$ between the plates,
the electric field will be of the same order. It compensates the root
coefficient in Eq. (15) and we get acceleration of about $1cm/s^2$, which is
perfectly measurable. Static magnetic fields create the same gravitational
effects as static electric fields \cite{fift,sixt,sevent}, so a field of $%
33.2T$ may induce in principle acceleration of about $10cm/s^2$. One needs
just two orders more to counter Earth's gravity. These effects are much
stronger than any other known general relativistic effects, including
gravitational waves and gravitomagnetism, which are currently under
intensive study. They may be produced in a finite region in a laboratory if
we learn how to create WMP fields. Therefore, in the next sections we
revisit these fields, putting the emphasis on their physical applicability.

\section{Weyl fields revisited}

Let us start with the axially-symmetric static metric 
\begin{equation}
ds^2=f\left( dx^0\right) ^2-f^{-1}\left[ e^{2k}\left( dr^2+dz^2\right)
+r^2d\varphi ^2\right] ,  \label{twenty}
\end{equation}
where $x^0=ct$, $x^1=\varphi ,$ $x^2=r,$ $x^3=z$ are cylindrical
coordinates, $f=e^{2u}$ and $u$ is the first, while $k$ is the second
gravitational potential. Both of them depend only on $r$ and $z$. Let $A_\mu
=\left( \bar \phi ,\bar \chi ,0,0\right) $ where $\bar \chi $ is the true
magnetic potential. Following Tauber \cite{eightt} we introduce the
auxiliary potential $\bar \lambda $%
\begin{equation}
\lambda _r=\frac fr\chi _z,\qquad \lambda _z=-\frac fr\chi _r,  \label{twone}
\end{equation}
so that 
\begin{equation}
F^{\varphi r}=\frac{fe^{-2k}}r\bar \lambda _z,\qquad F^{z\varphi }=\frac{%
fe^{-2k}}r\bar \lambda _r  \label{twtwo}
\end{equation}
describe the axial and the radial components of the magnetic field. For the
electric field one has 
\begin{equation}
E_r=F_{0r}=-\bar \phi _r,\qquad E_z=F_{0z}=-\bar \phi _z.  \label{twthree}
\end{equation}
The field equations read 
\begin{equation}
\Delta u=e^{-2u}\left( \phi _r^2+\phi _z^2+\lambda _r^2+\lambda _z^2\right) ,
\label{twfour}
\end{equation}
\begin{equation}
\Delta \phi =2\left( u_r\phi _r+u_z\phi _z\right) ,  \label{twfive}
\end{equation}
\begin{equation}
\Delta \lambda =2\left( u_r\lambda _r+u_z\lambda _z\right) ,  \label{twsix}
\end{equation}
\begin{equation}
\phi _r\lambda _z=\phi _z\lambda _r,  \label{twseven}
\end{equation}
\begin{equation}
\frac{k_r}r=u_r^2-u_z^2-e^{-2u}\left( \phi _r^2-\phi _z^2+\lambda
_r^2-\lambda _z^2\right) ,  \label{tweight}
\end{equation}
\begin{equation}
\frac{k_z}r=2u_ru_z-2e^{-2u}\left( \phi _r\phi _z+\lambda _r\lambda
_z\right) ,  \label{twnine}
\end{equation}
\begin{equation}
k_{rr}+k_{zz}=\Delta u-\left( u_r^2+u_z^2\right) ,  \label{thirty}
\end{equation}
where $\Delta =\partial _{rr}+\partial _{zz}+\partial _r/r$. We have used
the definition given in Eq. (10) and a similar one for $\lambda $. Using Eq.
(23) one can prove that $\lambda =\lambda \left( \phi \right) $ and the
dependence is linear \cite{fift,sixt}. This result holds also for a general
static metric \cite{sevent}. It is enough to engage just an electric field,
there being a trivial magnetovac analogue to every electrovac solution. Eqs.
(20-23) reduce to \cite{ninet} 
\begin{equation}
\Delta u=e^{-2u}\left( \phi _r^2+\phi _z^2\right) ,\qquad \Delta \phi
=2\left( u_r\phi _r+u_z\phi _z\right) ,  \label{thone}
\end{equation}
which determine $\phi $ and $f$. Eqs. (24,25) become 
\begin{equation}
\frac{k_r}r=u_r^2-u_z^2-e^{-2u}\left( \phi _r^2-\phi _z^2\right) ,\qquad 
\frac{k_z}r=2u_ru_z-2e^{-2u}\phi _r\phi _z  \label{thtwo}
\end{equation}
and determine $k$ by integration, while Eq.(26) holds identically and is
redundant. When $\phi =0$, Eq. (27) becomes the Laplace equation for $u$ and
Eq. (28) gives $k\left( u\right) $. These are the axially-symmetric vacuum
equations, also discovered by Weyl.

Now let us make the assumption that the gravitational and the electric
potential have the same equipotential surfaces, $f=f\left( \phi \right) $.
Eq. (27) yields 
\begin{equation}
\left( f_{\phi \phi }-2\right) \left( \phi _r^2+\phi _z^2\right) =0
\label{ththree}
\end{equation}
and that's how the quadratic relation (12) appears. Replacing it in Eq. (27)
one comes to an equation for $\phi $%
\begin{equation}
\Delta \phi =\frac{B+2\phi }{1+B\phi +\phi ^2}\left( \phi _r^2+\phi
_z^2\right) .  \label{thfour}
\end{equation}
We put for definiteness $B\geq 0$. Eqs. (12,30) with $B\leq 0$ are obtained
by changing the sign of $\phi $. The general solution of Eq. (30) is not
known. However, let us make one more assumption, that $\phi $ depends on $%
r,z $ through some function $\psi \left( r,z\right) $. Eq. (30) becomes 
\begin{equation}
\frac{\phi _{\psi \psi }}{\phi _\psi }-\frac{\left( B+2\phi \right) \phi
_\psi }{1+B\phi +\phi ^2}=-\frac{\Delta \psi }{\psi _r^2+\psi _z^2}.
\label{thfive}
\end{equation}
If $\psi $ satisfies the Laplace equation $\Delta \psi =0$, $\phi \left(
\psi ,B\right) $ is determined implicitly \cite{fourt} from 
\begin{equation}
\psi =\int \frac{d\phi }{1+B\phi +\phi ^2}.  \label{thsix}
\end{equation}
A very important equality follows 
\begin{equation}
\phi _i=f\psi _i,\qquad \bar \phi _i=f\left( \phi \right) \bar \psi _i,
\label{thseven}
\end{equation}
where $i=r,z$. Eq. (28) becomes 
\begin{equation}
k_r=\frac D4r\left( \psi _r^2-\psi _z^2\right) ,\qquad k_z=\frac D2r\psi
_r\psi _z,  \label{theight}
\end{equation}
where $D=B^2-4$. Obviously, $k\sim \kappa $ always, making it much smaller
than $u$, which is proportional to $\sqrt{\kappa }$.

Thus in Weyl electrovac solutions the harmonic master potential $\psi $
determines the electric and the gravitational fields like $u$ does this in
the vacuum case. One may go further and find a relation between $\psi $ and $%
u$, transforming Weyl electrovacs into Weyl vacuum solutions, although usual
transformations work the other way round \cite{three}. In particular
solutions $\phi $ is usually proportional to the charge, $\phi =q\tilde \phi 
$. Eq. (32) shows that $\psi =q\tilde \psi $ where $\tilde \psi $ is
harmonic and finite when the electric field is turned off by $q\rightarrow 0$%
. In this limit, when $B$ does not depend on $q$, we have $f\rightarrow 1$
from Eq. (12) and $k\rightarrow 0$ from Eq. (34). Trivial flat spacetime is
the result. However, if $B=\tilde B/q$ then $f\rightarrow 1+\tilde B\tilde 
\phi $, $f_i\rightarrow \tilde B\tilde \phi _i$ and from Eq. (33) it follows
that $u=\tilde B\tilde \psi /2$ and is harmonic. Eq. (34) then gives the
vacuum expression for $k$. Hence, we obtain a Weyl vacuum solution (induced
by some mass) with the same, up to a constant, harmonic function $\psi $. A
mass term has appeared out of the vanishing charge.

Let us go back to the electrovac problem. One can add a constant $\psi _0$
to $\psi $ in order to satisfy the conditions $\psi \rightarrow
0,f\rightarrow 1,\phi \rightarrow 0$ at infinity or when the electric field
is turned off. The integral in Eq. (32) can be evaluated analytically and
the dependence $\phi \left( \psi ,B\right) $ made explicit. There are three
cases, according to the sign of $D$. The simplest one is $D=0$ ($B=2$). Then 
$f$ becomes a perfect square and $\psi _0=-1$, 
\begin{equation}
\phi =-1-\frac 1{\psi +\psi _0}=\frac \psi {1-\psi },\qquad f=\left( 1-\psi
\right) ^{-2}.  \label{thnine}
\end{equation}

When $D<0$ Eq. (34) gives $-D<4$ and trigonometric functions appear 
\begin{equation}
\phi =-\frac B2+\frac{\sqrt{-D}}2\tan \frac{\sqrt{-D}}2\left( \psi +\psi
_0\right) ,\qquad \psi _0=\frac 2{\sqrt{-D}}\arctan \frac B{\sqrt{-D}},
\label{forty}
\end{equation}
\begin{equation}
f=-\frac D{4\cos ^2\frac{\sqrt{-D}}2\left( \psi +\psi _0\right) }=\left(
\cos \frac{\sqrt{-D}}2\psi -\frac B{\sqrt{-D}}\sin \frac{\sqrt{-D}}2\psi
\right) ^{-2}.  \label{foone}
\end{equation}
When $\psi _0\equiv 0$ these formulas coincide with the Bonnor's ones \cite
{ninet}.

Finally, when $D>0$ there exist two expressions for the integral, one as a
logarithm, the other in hyperbolic functions. They lead to 
\begin{equation}
\phi =-\frac B2-\frac{\sqrt{D}}2\coth \frac{\sqrt{D}}2\left( \psi +\psi
_0\right) =\frac{2\left( e^{\sqrt{D}\psi }-1\right) }{B+\sqrt{D}-\left( B-%
\sqrt{D}\right) e^{\sqrt{D}\psi }},  \label{fotwo}
\end{equation}
\begin{equation}
f=\frac D{4\sinh ^2\frac{\sqrt{D}}2\left( \psi +\psi _0\right) }=\left(
\cosh \frac{\sqrt{D}}2\psi -\frac B{\sqrt{D}}\sinh \frac{\sqrt{D}}2\psi
\right) ^{-2},  \label{fothree}
\end{equation}
\begin{equation}
e^{\sqrt{D}\psi _0}=\frac{B-\sqrt{D}}{B+\sqrt{D}}.  \label{fofour}
\end{equation}

According to Bonnor (who does not introduce $\psi _0$) the expressions for $%
f $ and $\phi $ in the case $D>0$ are obtained from Eq. (36) by continuation
of $\sqrt{-D}$ to imaginary values and consequently $\tan \rightarrow \tanh $%
, $\cos \rightarrow \cosh $. This, however, holds when $\left( 2\phi
+B\right) ^2<D$ , i.e., $4f<0$, which is unphysical. In the physical case we
must also do the replacement $\tanh \rightarrow \coth $, $\cosh \rightarrow
\sinh $. The above discussion shows that the point $B=2$ has a privileged
position, unlike the point $B=0$.

The Weyl solutions were derived with two assumptions imposed on the system: $%
f=f\left( \phi \right) $; $\phi =\phi \left( \psi \right) $, $\Delta \psi =0$%
. They are particular solutions of Eqs. (27,28). However, when the symmetry
is stronger than axial and the fields depend on just one coordinate $x$ (not
necessarily cylindrical, but a function of $r,z$), they comprise the general
solution. For then $f\left( x\right) $ and $\phi \left( x\right) $ obviously
are functionally related and one can always find $X\left( x\right) $ so that 
$\Delta X\left( x\right) =0$. Taking $\psi =X\left( x\right) $ and
expressing $f$ and $\phi $ as functions of $X$, leads inevitably to the Weyl
solutions. In the case of plane symmetry $x=z$, $X=x$. Cylindrical symmetry
gives $x=r$, $X=\ln x$. Spherical symmetry has $x=\sqrt{r^2+z^2}$, $X=1/x$.

We have described the advantages of WMP solutions in inducing a powerful
gravitational force. Some natural questions appear: is it possible to create
such fields in a laboratory? What kind of charged sources should we take?
Point \cite{twenty} and line \cite{twone} sources lead to singularities and
other problems. Therefore, we take a charged, closed, rotationally-symmetric
surface with invariant density of the surface charge $\sigma _s$. The
electrostatic theorem of Gauss has a generalization in general relativity 
\cite{thirt,twtwo}. Integrating the r.h.s. of Eq. (5) one obtains the total
charge contained in some volume 
\begin{equation}
e=\frac 1c\int J^0\sqrt{-g}d^3S=\int \sigma _3d^3S,\quad \sigma _3=\sigma 
\sqrt{-g^{\left( 3\right) }},  \label{fofive}
\end{equation}
where $\sigma _3$ is the three-dimensional invariant density and $g^{\left(
3\right) }$ is the determinant of the space part of the metric. When the
charge is attached to a surface, one should use the surface charge density $%
\sigma _s$ instead. Integration of the l.h.s. of Eq. (5) leads to a relation
between $e$ and the electric flux through a closed surface $S$, encompassing
the charged volume

\begin{equation}
4\pi e=\int_S\left[ F^{01}\frac{\partial \left( x_2,x_3\right) }{\partial
\left( u,v\right) }+F^{02}\frac{\partial \left( x_3,x_1\right) }{\partial
\left( u,v\right) }+F^{03}\frac{\partial \left( x_1,x_2\right) }{\partial
\left( u,v\right) }\right] \sqrt{-g}dudv.  \label{fosix}
\end{equation}
For Weyl solutions 
\begin{equation}
F^{0i}\sqrt{-g}=-r\bar \psi _i,  \label{foseven}
\end{equation}
which is the result for flat spacetime. Hence, Eq. (42) becomes the Gauss
theorem in classical electrostatics, but with $\phi $ replaced by $\psi $,
which satisfies the Laplace equation. This fact was already stressed by
Bonnor \cite{five,ninet} but in view of its importance we have discussed it
again. Following a well-known procedure, we obtain a boundary condition on $%
S $ for the jump of the normal component $\bar \psi _n$: 
\begin{equation}
-\bar \psi _n|_{-}^{+}=4\pi \bar \sigma _s.  \label{foeight}
\end{equation}
When $\psi $ is given on $S$, there are two well-defined Dirichlet boundary
problems and it may be continued as a harmonic function inside and outside $%
S $. If $\alpha \leq \psi _s\leq \beta $, these inequalities hold for $\psi $
throughout space and it will be regular. Then $f$, $k$ and $\phi $ are found
from $\psi $ in a manner already explained. The jump of $\psi _n$ at $S$
determines the source $\sigma _s$. The inverse is also true. When $\sigma _s$
is given, there is a unique global $\psi $, satisfying Eq. (44).
Consequently, for any distribution of charges on $S$ one can find the
electric and gravitational fields they induce. The same can be done when $S$
is infinite and (or) not closed, but singularities may creep into the
solutions.

Finally, by replacing Eq. (33) into Eq. (13) and expressing $f=f\left( \psi
\right) $ we obtain 
\begin{equation}
g_i=\frac 12\sqrt{G\left( D+4f\right) }\bar \psi _i=\sqrt{Gf}\bar \psi
_i|_{B=2}.  \label{fonine}
\end{equation}
For realistic EM-fields $f$ is very close to one and this Maxwellian effect
is the only one to be observed. Typical Einsteinian effects like light
bending, gravitational redshift, time delay, changes in lengths are not
enhanced by $c^2$ and are negligible.

Some questions immediately arise. Why do we get a Weyl solution for an
arbitrary $\sigma _s$ when Weyl fields are not the most general solutions of
Eq. (27)? Why $\phi ,f,k$ depend not alone on $\psi $ but also on the
constant $B$? How is its value determined, can we increase it, to enhance
the effect of root gravity? In order to answer them we must return to the
starting point, Eq.(3). Even when the EM-field is absent, Eq. (3) still has
a number of non-trivial vacuum solutions, including e.g. gravitational
waves. The situation is similar to classical electrodynamics without
sources. Non-trivial solutions exist, but have to be time-dependent. These
are the well-known electromagnetic waves. General relativity is a highly
non-linear theory and vacuum solutions exist also in the static case. Their
sources are well-hidden masses and even today there is a gap between the
mathematical derivation of solutions \cite{three} and their physical
interpretation \cite{twthree}. When EM-fields are turned on, these parasitic
masses do not disappear and obscure the pure effect of electromagnetism on
gravity. Let us try to get rid of them, step by step. First of all, the
metric should inherit the symmetry of EM-fields. Let us confine again
ourselves to axial symmetry. There are a lot of generation techniques, which
produce non-Weyl solutions of Eq. (31). Most of the methods (see Ref. \cite
{three}, Ch.34) start from the reformulation of Eq. (27) in terms of the
Ernst potential $E=f-\phi ^2$ \cite{twseven}, which is real in the absence
of rotation, 
\begin{equation}
f\Delta E=\nabla f\nabla E,\quad f\Delta \phi =\nabla f\nabla \phi .
\label{fifty}
\end{equation}
The general solution of the Ernst equations can be found when the behaviour
of $E\left( z\right) $ and $f\left( z\right) $ on the axis is given. It
determines the multipole structure and is useful in astrophysics for
modelling the gravitational field of stars. The presence of masses is
welcomed, since they give the most substantial gravitational effect,
followed by rotation (it can be incorporated into the formalism), magnetic
fields and electric charge at the last place. The Ernst equation is much
more difficult than the Laplace one and Dirichlet boundary value problems
for it were discussed only recently \cite{tweight}. On the other side, a
harmonic function may be easily restored from its values on the axis $\psi
\left( z\right) $ \cite{twnine,thirty} 
\begin{equation}
\psi \left( r,z\right) =\frac 1\pi \int_0^\pi \psi \left( z+ir\cos \theta
\right) d\theta .  \label{fione}
\end{equation}
This real expression was used in general relativity for axially-symmetric
static vacuum solutions \cite{thone}, but it is not difficult to adapt it to
Weyl electrovacs too. In the general solution $f\left( z\right) $ is not
correlated with $\phi \left( z\right) $ and can be arbitrary, even when $%
\phi \left( z\right) $ vanishes, signalling the presence of masses. In the
Weyl solution the metric goes to flat Minkowski spacetime when $\psi \left(
z\right) $ vanishes. Thus the general solution expands over the Weyl one by
the addition of masses. Probably the same is true for the solutions of Eq.
(30), which also contain root gravity terms, since Eq. (12) is satisfied.
The reason is that Weyl solutions form a complete system, covering the
effect of any charge distribution and more general solutions can include the
only other source of gravitation. More precisely, Weyl fields form an
overcomplete system due to $B$ which is not fixed by $\sigma _s$. In the
case of plane, spherical, spheroidal or cylindrical symmetry they comprise
the most general solution of the Ernst equation (46). It is quite improbable
that the unwanted masses should disappear exactly in these cases, so the
only way to show their presence is through the value of $B$.

A logical step is to accept that $\psi $ plays the role of $\phi $ in any
situation in electrogravity, not only for charged surfaces. Hence, when the
gravitation created by charges is taken into account, it seems that Weyl
fields generalize the solution to classical electrostatic problems. The
physical electric field is found with the help of Eq. (33) 
\begin{equation}
E_{\left( i\right) }=-\left( g_{00}g_{ii}\right) ^{-1/2}\bar \phi _i=-fe^{-k}%
\bar \psi _i.  \label{fitwo}
\end{equation}
Now, since $f,k$ are extremely close to one and zero respectively, one can
do a perturbation theory around the exact Weyl solutions and set $E_{\left(
i\right) }\approx -\bar \psi _i$. In the present case WMP fields are similar
to instantons, monopoles, solitons and other non-perturbative exact
solutions in quantum field theory. With high precision all electrostatic
formulae hold also in the Einstein-Maxwell theory. The only new effect is
the appearance of an electromagnetically induced gravitational acceleration,
which reads from Eq. (45), again with high precision, 
\begin{equation}
g_i=-\frac B2\sqrt{G}E_i.  \label{fithree}
\end{equation}
We shall give arguments in the following that $B=2$, when unbiased by
parasitic masses. Therefore, as already explained, measurable $g_i$ are
present from the available today electric and magnetic fields. One can reach 
$g_e$ when $E_i=1.14\times 10^9V/cm=3.8\times 10^6CGS$ or $%
H_i=380T=3.8\times 10^6G.$ The lines of acceleration follow the electric
field lines. Test particles will stay in equilibrium if they are charged and
the relation between their mass $m$ and charge $e$ is 
\begin{equation}
\left| e\right| =\sqrt{G}m.  \label{fifour}
\end{equation}

Root gravity has some peculiar features. Changing the direction of $E_i$ one
changes the direction of $g_i$ and when it points upwards with respect to
the Earth's surface one has ''anti-gravity''. This is true because in our
perturbation theory accelerations from electric fields and masses like the
Earth or laboratory masses are added as usual vectors. The exact Weyl
solution is necessary to clarify the gravitational induction in a laboratory
set-up of finite size in space, where $E_i$ is present. Although we have
used a long-range interaction to induce another long-range interaction, in
reality static EM-fields are always confined. Eq. (49) shows that putting a
Faraday cage on $E_i$ confines $g_i$ too. It is understandable that when
non-mass sources of gravitation are applied, the appearance of gravitational
monopole terms (usually considered as mass terms) is not obligatory. Their
existence is usually based on the Whittaker's theorem \cite{twtwo}, which
demonstrates the influence of some combination of the $T_{\mu \nu }$
components (called gravitational mass) upon the gravitational acceleration.
The relation is given in terms of surface and volume integrals, appearing
when the time-time component of the Einstein equations (1) is integrated. In
this way it is not something separate and additional to them, but a
consequence that can't contradict the conclusions following from them.
Concretely, for Weyl fields this theorem is just the Gauss theorem for $%
grad\;u$. In the case of electromagnetic sources the ''gravitational mass''
is in fact some kind of ''energy'', inducing gravitational acceleration not
necessarily with a monopole term. We avoid arguments based on the energy of
the gravitational field because its density is not a tensor and there are at
least five energy-momentum complexes \cite{ththree}, each with its own
merits. Of course, some small mass term in the acceleration will always
exist, due to the mass of the surface $S$. However, it will be of usual
perturbative nature, many orders of magnitude smaller than root gravity.

Let us turn now to the case of magnetostatics. As was mentioned before, the
analogue of $\phi $ is $\lambda $. It should be replaced in Eqs.
(12,13,27,33,35,36,38). The analogue of Eq. (42) vanishes because there
are no magnetic charges. One should take a closed surface with a surface
current. In an axially-symmetric problem it has just one component, $%
J_\varphi $. A Weyl magnetostatic solution was given for the first time by
Papapetrou in 1947 \cite{twelve}. The analogy with electrostatics was
investigated by Bonnor \cite{five,ninet} who showed that $\psi $ is
equivalent to the scalar magnetic potential. Then Eq. (44) should give the
jump of the tangential to the surface component $H_t$ which follows
classically from $\bar \psi $ 
\begin{equation}
H_t|_{-}^{+}=\frac{4\pi }cJ_\varphi  \label{fifive}
\end{equation}
and is perpendicular to $J_\varphi $. Eqs. (48,49) still hold with $%
E_{\left( i\right) }\rightarrow H_{\left( i\right) }$, so that the
gravitational effects of static magnetic fields mirror those in
electrostatics. One can also introduce a vector potential, corresponding to $%
\psi $, which is more convenient in classical magnetostatics. However, in
order to find the metric, the scalar potential $\psi $ for each classical
problem should be calculated too. The fields mainly of linear sources, like
a current loop \cite{twone}, and disks \cite{thfour,thfive} have been
examined without discussing the presence of root gravity.

\section{Connections and generalizations}

In the general stationary case the interval reads 
\begin{equation}
ds^2=f\left( dx^0+\omega _adx^a\right) ^2-f^{-1}\gamma _{ab}dx^adx^b,
\label{fisix}
\end{equation}
where $\omega _a$ is the gravitomagnetic potential ($a=1,2,3$) and $\gamma
_{ab}$ is the three-dimensional metric. In the general static case $\omega
_a=0$ and in its electrostatic subcase the Einstein-Maxwell equations read 
\begin{equation}
\Delta u=e^{-2u}\nabla \phi \nabla \phi ,\quad \Delta \phi =2\nabla u\nabla
\phi ,  \label{fiseven}
\end{equation}
\begin{equation}
R_{ab}^{\left( 3\right) }=2u_au_b-2e^{-2u}\phi _a\phi _b,  \label{fieight}
\end{equation}
where $F_{0a}=-\bar \phi _a$. In magnetostatics $\phi \rightarrow \lambda $,
the latter being defined by 
\begin{equation}
F^{ab}=\left( -g\right) ^{-1/2}\varepsilon ^{abc}\bar \lambda _c.
\label{finine}
\end{equation}
This equation generalizes Eq. (18). The gradients, the Laplacian and the
three-dimensional Ricci tensor are with respect to the metric $\gamma _{ab}$%
. Eqs. (53,54) generalize Eqs. (27,28). However, $\gamma _{ab}$ also enters
Eq. (53), making all equations interconnected and the system very difficult
to deal with. In the special case when Eq. (12) holds with $B=2$, $\gamma
_{ab}$ becomes flat and Eq. (53) can be solved, since it decouples from Eq.
(54). The result is Eq. (35) with a harmonic $\psi \left( \varphi
,r,z\right) $. Usually one takes $1-\psi =U$ and $\phi =U^{-1}$, obtaining
the already mentioned Majumdar-Papapetrou solutions from 1947 \cite
{eleven,twelve}, which are conformastatic. Some years later Ehlers \cite
{thnine,forty} gave transformations to derive such fields from vacuum ones.
Similar transformations were given by Bonnor \cite{foone} and with the help
of the TWS method \cite{fotwo,fothree}. The latter was applied to the RN
solution \cite{fothree,fofour}.

When there is no space symmetry present, the simplest harmonic function in
cartesian coordinates is 
\begin{equation}
U\left( x,y,z\right) =1+\sum\limits_i\frac{Gm_i}{c^2r_i},\quad r_i=\left[
\left( x-x_i\right) ^2+\left( y-y_i\right) ^2+\left( z-z_i\right) ^2\right]
^{1/2}.  \label{sixty}
\end{equation}

It can be shown that the sources are point monopoles with masses $m_i$ and
charges $e_i$ \cite{thirt,twenty}, connected by Eq. (50), which is true also
in the Newtonian theory. It ensures the equilibrium between the electric and
gravitational forces among the sources. Such multi-black-hole solutions
satisfy certain uniqueness theorems and possess axially-symmetric reduction
with the points staying on the $z$-axis \cite{foseven}. They have been
generalized to expanding cosmological solutions \cite{foeight,fonine}.

Charged dust solutions with mass density $\mu $ and charge density $\sigma $
should be mentioned too. For this purpose one must add to the r.h.s. of Eq.
(3) $T^{\mu \nu }=\mu c^2u^\mu u^\nu $ and introduce the current in Eq. (5).
In the general static case one can impose the Newtonian equilibrium
condition 
\begin{equation}
\pm \sigma =\sqrt{G}\mu ,  \label{sione}
\end{equation}
which is the density analogue of Eq. (50), but here it holds for the sources
of the field, not for a test-particle. Then one gets \cite{fione} 
\begin{equation}
f=\left( C+\phi \right) ^2,  \label{sitwo}
\end{equation}
where $C$ is some constant. Like in the multi-black-hole case one puts $\phi
=U^{-1}$, $f=U^{-2}$, but $U$ is not harmonic; it satisfies the non-linear
equation \cite{eleven,fione} 
\begin{equation}
\Delta U=\pm \frac{4\pi \sqrt{G}}{c^2}\sigma U^3.  \label{sithree}
\end{equation}

Charged dust clouds of spherical or spheroidal shape and obeying Eq. (59)
have been studied extensively in an astrophysical context \cite
{fifour,fieight}. They have a number of interesting properties when compared
to usual stars: their mass and radius may be arbitrary, very large redshifts
are attainable, their exteriors can be made arbitrarily near to the exterior
of extreme charged black holes. In the spherical case the average density
can be arbitrarily large, while for any given mass the surface area can be
arbitrarily small. When their radius shrinks to zero, many of their
characteristics remain finite and non-trivial.

Some more general solutions of Eq. (59) have been given too \cite
{finine,sixty}. They include the case of constant $\mu $ with $U$ given in
terms of a Jacobi elliptic function. The idea that $\mu $ may be
concentrated on surfaces was also discussed \cite{sione}. Thin dust shells
of spherical, cylindrical or plane shape were given as examples. Finally, a
magnetostatic dust solution with functional dependence between $f$ and $%
A_\varphi $ is also known \cite{sitwo}.

The charge density required to satisfy Eq. (57) is quite small. It is
sufficient that in a sphere of neutral hydrogen one atom in about $10^{18}$
had lost its electron \cite{fifour}. However, all charged dust models have
one essential flaw: the equilibrium is very delicate and unstable and a
slight change in $\sigma $ would cause the cloud to expand or contract. On
the other hand, the models discussed in the present paper do not depend on
equilibrium conditions. They require just the creation of strong enough
electric or magnetic fields. Of course, if the charged cloud is divided into
parts with positive and negative particles (or ions), this can also induce
EM-fields and root gravity.

The appearance of the WMP relation (12) was studied also when the dust is
pressurized, i.e., in the case of charged perfect fluids and a gravitational
model for the electron was put forth \cite{sifive}. In both cases many
different relations between $f$ and $\phi $ are possible. Spherical charged
perfect fluid interior solutions are reviewed in Ref. \cite{siseven}.

\section{Electrostatic examples}

In this section two simple experimental set-ups are given where the effects
of root gravity may be detected. They involve plane, spherical, and
spheroidal symmetry. In these cases Weyl fields represent the most general
electrovac solution and there is no ambiguity as to their appearance.

\subsection{Plane symmetry}

The study of plane-symmetric electrovac metrics goes back to 1926 \cite
{eifive}, but for a long time their Weyl nature remained unrecognized. A
plane-symmetric example of a WMP field was given first by Papapetrou \cite
{twelve}. Later Bonnor studied fields with $\phi =\phi \left( z\right) $,
which include also some non-Weyl solutions \cite{ninet}. Let us discuss the
gravitational field of a uniformly charged plane with charge density $\sigma 
$. The classic Maxwell potential (which becomes the master-potential) has
both plane and mirror symmetry 
\begin{equation}
\psi =-2\pi \sigma \left| z\right| \equiv q\left| z\right| .  \label{seeight}
\end{equation}
It vanishes at the plane $z=0$ and goes to infinity when $\left| z\right|
\rightarrow \infty $. In the case $D=0$ we should replace Eq. (60) into Eq.
(35). Kar has proposed a coordinate system where $\phi $ becomes harmonic, $%
\phi =q\left| z^{\prime }\right| $. It gives a constant electric field and
further deepens the analogy with classical electrostatics. Then the interval
becomes 
\begin{equation}
ds^2=f\left( dx^0\right) ^2-f^{-1}\left( dr^2+r^2d\varphi ^2\right)
-f^{-3}\left( dz^{\prime }\right) ^2,  \label{senine}
\end{equation}
where $f=\left( 1+q\left| z^{\prime }\right| \right) ^2$. When $D\neq 0$ one
sees from Eq. (34) that $k$ depends on $r$ and in fact 
\begin{equation}
k=-\frac D8q^2r^2.  \label{eighty}
\end{equation}
The metric does not inherit the symmetry of its source. Now the Kar's gauge
gives $\phi \left( z\right) =-B/2+q\left| z^{\prime }\right| +\alpha $.
Requiring flatness at $z^{\prime }=0$ \cite{ninet} one obtains $\alpha =B/2$
and Eqs. (36,37) yield 
\begin{equation}
ds^2=f\left( dx^0\right) ^2-e^{q^2\left( 1-\alpha ^2\right) r^2}\left(
f^{-1}dr^2+f^{-3}dz^{\prime 2}\right) -f^{-1}r^2d\varphi ^2,  \label{eione}
\end{equation}
where $f=1+2q\alpha \left| z^{\prime }\right| +q^2z^{\prime 2}$. The
non-inheritance is obvious again. The $D=0$ case is obtained when $\alpha
=\pm 1$ and is the only one with plane symmetry. This is a strong argument
that $B=2$. Other values of $B$ (including $B=0$) introduce, in addition,
the parameter $\alpha $, which does not originate from the electric field,
unlike the charge parameter $q$. Probably it is invoked by hidden mass
sources. As for the condition $f\left( z=0\right) =1$, its necessity will
become clear in the following. In conclusion, the only plane-symmetric Weyl
metric is given by Eq. (61). Another example of non-inheritance was given in
Ref. \cite{eisix}, where the metric is not rotationally invariant.

In order to obtain a regular global solution, one must satisfy the junction
conditions \cite{eieight} at the plane. The metric is continuous there. The
extrinsic curvature reads 
\begin{equation}
K_{aa}=\frac 12\left| g_{zz}\right| ^{-1/2}\left( g_{aa}\right) _z,
\label{eitwo}
\end{equation}
where $a=x^0,\varphi ,r$ and there is no summation. Its eventual jump at the
plane determines the required energy-momentum tensor $S_{ab}$ of the massive
surface layer 
\begin{equation}
\kappa S_{\;b}^a=\gamma _{\;b}^a-\delta _{\;b}^a\gamma ,\quad \gamma
_{\;a}^a=K_{\;a}^a|_{-}^{+},\quad \gamma =\sum\limits_a\gamma _{\;a}^a.
\label{eithree}
\end{equation}
For the Weyl form of the axially-symmetric metric one has 
\begin{equation}
\kappa S_{\;0}^0=e^{u-k}\left( 2u_z-k_z\right) |_{-}^{+},\quad
\label{eifour}
\end{equation}

\begin{equation}
\kappa S_{\;\varphi }^\varphi =-e^{u-k}k_z|_{-}^{+},\quad \kappa S_{\;r}^r=0.
\label{eifive}
\end{equation}
In our case $k_z=0$ and even $k=0$. At first sight, the jump in the
acceleration $u_z$ requires the introduction of mass on the charged plane.
However, its true cause is the electric field present in the space around
the plane and induced by the charge distribution on it. This is seen when
one makes use of Eqs. (12,33) 
\begin{equation}
u_z=\left( 1+\phi \right) \psi _z.  \label{eisix}
\end{equation}
Obviously, the jump in $u_z$ at $z=0$ is due to the jump in the
master-potential $\psi _z$ and consequently to the presence of charge. The
charged surface layer of the plane causes the jump in the extrinsic
curvature too. For Weyl fields the usual procedure of introducing mass and
pressures on the surface represents an attempt to model the influence of the
electric field upon the metric by traditional mass sources \cite
{thfour,thfive,einine}. When $\psi $ from Eq. (60) is replaced in $g_i$, the
plane will be attractive for one sign of $q$, as if there was positive mass
on it, but for the other sign of $q$ it will be repulsive as if it were made
of exotic negative mass, which breaks the energy conditions. There is no
paradox because the true creator of these effects is the energy-momentum
tensor of the electric field, which always satisfies all three energy
conditions.

Now let us put a second plane at $z=d$, charged in the opposite way. The
electric field is confined between the two planes and $\psi $ reads 
\begin{equation}
\psi =\frac{\psi _2}dz,\quad E_z=-f\bar \psi _z.  \label{eiseven}
\end{equation}
Here $\psi _2$ is the potential of the second plane ($\psi _1=0$). It is
related to the charge density by 
\begin{equation}
\psi _2=2\pi \sigma \frac d\varepsilon ,  \label{eieight}
\end{equation}
where $\varepsilon $ is the dielectric constant. Up to now we have
considered the case $\varepsilon =1$. More generally, $\varepsilon $ enters
Eq. (3) because the energy $T_{00}\sim \varepsilon E_i^2$ and consequently $%
T_{\mu \nu }\rightarrow \varepsilon T_{\mu \nu }$. When $\phi $ absorbs the
constants in Eq. (10) it will pick also $\sqrt{\varepsilon }$. The same is
true for $\psi $. The acceleration formula (45) becomes in a dielectric
medium 
\begin{equation}
g_z=\sqrt{G\varepsilon f}\frac{\bar \psi _2}d\approx 2.58\times 10^{-4}\frac{%
\sqrt{\varepsilon }}d\bar \psi _2.  \label{einine}
\end{equation}

Taking finite disks instead of planes one obtains the usual capacitor. The
field around its centre will be plane-symmetric, while at the rim it will
depend on $r$ too. This may be diminished by careful electric shielding of
the capacitor. Outside there will be a vacuum flat metric, which joins
smoothly the interior due to the conditions $f=const$ on the plates. One
comes to the conclusion that the capacitor will be subjected to practically
constant gravitational force $F_g$ in the $z$-direction, 
\begin{equation}
F_g=\sqrt{G\varepsilon }\frac Md\bar \psi _2=\sqrt{G\varepsilon }\mu S\bar 
\psi _2,  \label{ninety}
\end{equation}
where $M$ is the mass of the dielectric, $\mu $ is its mass density and $S$
is the area of the plate. This force is very different from the electric
force, trying to bring the plates together 
\begin{equation}
F_E=\frac{\varepsilon ^2S\bar \psi _2^2}{2\pi d^2}.  \label{nione}
\end{equation}
The latter is neutralized by the mechanical construction of the capacitor.
If it is hanging freely, the effect of $F_g$ may be tested experimentally.
To increase the acceleration it is advantageous to make $d$ small (typically 
$0.1cm\leq d\leq 1cm$), to raise the potential difference $\psi _2$ between
the plates up to $2\times 10^4CGS$ and to take a material with high $%
\varepsilon $. Some examples are gases ($\varepsilon \approx 1$), quartz ($%
4.5$), glycerine ($56.2$), water, electric ceramics ($81$), rutile ($TiO_2$)
with $\varepsilon =170$. Ferroelectrics are even better if one can cope with
their hysteresis and tendency for saturation in strong fields. Barium
titanate ($BaTiO_3$) and many others have $\varepsilon $ in the range of $%
10^4$. Thus $\sqrt{\varepsilon }/d$ may reach in principle $10^3$ and the
maximum acceleration $g_{z,\max }=5.2g_e$ is more than enough to counter
Earth's gravity. In a more modest attempt one can take $\bar \psi _2=100kV$
and $\sqrt{\varepsilon }/d=10^2$ to get about one percent of $g_e$.
Curiously, the same factors $\varepsilon ,d,\bar \psi _2$ are even more
important in $F_E$, which is typically much bigger than $F_g$.

As we shall see, root gravity effect is strongest in this first capacitor
example and, more generally, when the electric (magnetic) field lines are
parallel. The gravitational acceleration is not strictly constant, having an
unobservable dependence on $z$. One cannot obtain root gravity terms by
studying fields with constant acceleration \cite{ninety}. Earth's field also
shares these ''shortcomings'' of artificial gravity: the acceleration's
directions are not parallel but meet at the planet's centre and its
magnitude decreases with height.

In the traditional approach to the charged plane its metric depends directly
on $t$ and $z$, bypassing the axially-symmetric step, see \cite{three},
p234. The Weyl nature is then completely obscured. Some plane metrics are
induced by null EM-fields, like the special pp-wave given by Eq. (15.18)
from \cite{three} or the Robinson-Trautman solution, given by Eq. (28.43)
from the same reference. They are non-static because null EM-fields are
incompatible with static metrics, \cite{three}, Theorem 18.4, \cite{nione}.
We also do not discuss the non-null homogenous and uniquely conformally flat
Bertotti-Robinson solution, \cite{three} Sec.12.3. All other plane-symmetric
solutions with non-null EM fields have been found by Letelier and Tabensky 
\cite{nitwo}. The metric is either static or spatially homogenous. We are
interested in the static branch. It has been reviewed in Ref. \cite{nithree}%
, however, the electric field was not discussed and the ties with the Weyl
fields remained unelucidated. Let us clarify this issue now.

The metric reads in cylindrical coordinates 
\begin{equation}
ds^2=Kc^2dT^2-N\left( dR^2+R^2d\Phi ^2\right) -PdZ^2,  \label{nitwo}
\end{equation}
where $K,N,P$ depend on $Z$ and there is one relation between them. It
allows to express $K$ and $P$ as functions of $N$ 
\begin{equation}
K=\frac 1N\left( \beta _1+\beta _2\sqrt{N}\right) ,\quad P=\frac{N_z^2}{4\mu
^2}\left( \beta _1+\beta _2\sqrt{N}\right) ^{-1},  \label{nithree}
\end{equation}
\begin{equation}
\beta _1=\frac{\eta ^2}{4\mu ^2},\quad \beta _2=1-\beta _1,  \label{nifour}
\end{equation}
where $\eta $ and $\mu $ are constants related to the charge and mass
density on the plane. For the electric field we get from Eq. (5) 
\begin{equation}
E_z=-\phi _z=-\frac{\eta \sqrt{KP}}{2N},\quad \phi =\frac \eta {2\mu }\left( 
\frac 1{\sqrt{N}}-1\right) .  \label{nifive}
\end{equation}
We have chosen the boundary conditions $\phi \left( 0\right) =0$, $N\left(
0\right) =1$. It is easily seen that 
\begin{equation}
K=1+B_0\phi +\phi ^2,\quad N=\left( 1+\frac{2\mu }\eta \phi \right) ^{-2},
\label{nisix}
\end{equation}
\begin{equation}
B_0=\frac{2\mu }\eta +\frac \eta {2\mu }=\frac{2-\beta _2}{\sqrt{1-\beta _2}}%
.  \label{niseven}
\end{equation}
Thus $K$, the analogue of $f$, obeys Eq. (12) and root gravity terms are
present. When $\phi $ is turned off by $\eta \rightarrow 0$ ($\beta
_1\rightarrow 0$), $B_0$ does not stay constant. It increases to infinity
instead, so that $B_0\phi $ remains finite. This triggers the ''mass out of
charge'' mechanism described in Sec.III and we end with the vacuum solution
for a massive plane with $K=N^{-1/2}$ \cite{nifour}. Hence, the mass is not
entirely of electromagnetic origin and the parameter $\mu $ is independent
from $\eta $ in general. Another limit is $\beta _2\rightarrow 0$. This was
explored first by McVittie \cite{nifive}. Then $\beta _1=1$, $\eta =\pm 2\mu 
$, $B_0=\pm 2$, $K=1/N$.

One can absorb $N$ in Eq. (74) by the change $Z\rightarrow N$. However, it
is better to determine $N$ from the relation between $K,P$ and $N$ which
specifies the coordinate system. The conformastatic spacetimes studied by
Weyl \cite{fourt}, Papapetrou \cite{twelve} and Bonnor \cite{ninet} have $%
N=P $. Kar \cite{eifive} used two gauges, $KP=1$ and $KP=N^2$ in his
pioneering work. The second gives constant $E_z$, as seen from Eq. (77).
McVittie \cite{nifive} worked in the gauge $KP=N$. Patnaick \cite{nisix}
attacked the problem with $K=P$, which is the Taub's gauge in the vacuum
case \cite{niseven}. The same gauge was utilized by Letelier and Tabensky 
\cite{nitwo}. Unfortunately, the equations cannot be integrated explicitly
in this gauge except for the McVittie limit.

In order to make comparison between the axially-symmetric approach and the
traditional plane-symmetric approach one should use the gauge $N=P$. This
differential equation for $N$ is easily solved and yields 
\begin{equation}
N=P=\left( 1-\mu Z+\frac{\beta _2\mu ^2}4Z^2\right) ^2,\quad K=\frac 1N%
\left( 1-\frac{\beta _2\mu }2Z\right) ^2,  \label{nieight}
\end{equation}
\begin{equation}
\phi _z=\frac \eta {2N}\left( 1-\frac{\beta _2\mu }2Z\right) ,\quad \phi =%
\frac \eta {2\mu }\left[ \left( 1-\mu Z+\frac{\beta _2\mu ^2}4Z^2\right)
^{-1/2}-1\right] .  \label{ninine}
\end{equation}
These expressions coincide with the $B=2$ case, Eqs. (35,60) only in the
McVittie limit where $\beta _2=0$ and consequently $q=\mu =\eta /2$, $\eta
=-4\pi \sigma $. In this limit the fields in Kar's gauge \cite
{eifive,nithree} coincide with Eq. (63) for $\alpha =1$ after the
identification $Z=z^{\prime }$. The original McVittie's metric is obtained
by passing to $e^{qz^{\prime \prime }}=qz^{\prime }+1$ 
\begin{equation}
ds^2=e^{2qz^{\prime \prime }}\left( dx^0\right) ^2-e^{-2qz^{\prime \prime
}}\left( dx^2+dy^2\right) -e^{-4qz^{\prime \prime }}\left( dz^{\prime \prime
}\right) ^2.  \label{hundred}
\end{equation}

In conclusion, the purely electric effect upon the metric of a charged plane
is given by the McVittie solution, while the deviation of $\beta _2$ from
zero ($\mu $ from $\eta /2$) signals the presence of mass in addition to the
charge. In this way one can increase (very inefficiently) $B_0$ and the root
gravity acceleration.

It can be seen \cite{einine} that in the vacuum case a mass surface layer
has to be introduced in order to explain the mass term, appearing in the
metric. However, the electrostatic field creates itself a monopole term,
indistinguishable from the mass term (electromagnetic mass) and, hence, a
jump in the extrinsic curvature. Therefore, it is not necessary to introduce
compensating masses and pressures on the charged plane, in confirmation of
what was established above.

\subsection{Spherical symmetry}

Let us study now the gravitational effect of electric fields with spherical
symmetry. The master potential is 
\begin{equation}
\psi =\frac q\rho ,\quad \rho ^2=r^2+z^2.  \label{hsix}
\end{equation}
This is the exterior solution for a metal conductive sphere of radius $\rho
_1$, charged to a potential $\psi _1=q/\rho _1$. Inside the sphere $\psi =0$
and the spacetime is flat. Outside we obtain the charged \cite
{nieight,ninine} Curzon \cite{hundred} solution. Eq. (34) gives 
\begin{equation}
k=-\frac{Dq^2r^2}{8\rho ^4}.  \label{hseven}
\end{equation}
Like in the plane-symmetric case the metric does not inherit the spherical
symmetry of $\psi $ unless $D=0$ ($B=2$). This corresponds to the critically
charged Curzon metric. Eq. (35) yields 
\begin{equation}
\phi =\frac q{\rho -q},\quad f=\left( 1-\frac{2q}\rho +\frac{q^2}{\rho ^2}%
\right) ^{-1}.  \label{height}
\end{equation}
The charge in CGS units $\bar q$ is connected to $q$ by $q=\frac{\sqrt{G}}{%
c^2}\bar q$. Eq. (45) gives for the acceleration 
\begin{equation}
g_\rho =-\sqrt{Gf}\frac{\bar \psi _1\rho _1}{\rho ^2}.  \label{hnine}
\end{equation}
It has a maximum at the sphere and utilizing our maximum electric potential
we obtain $\left| g_{\rho ,\max }\right| =5.06/\rho _1$. When $\rho _1=10cm$%
, one gets $0.5cm/s^2$. Formula (64) for the extrinsic curvature holds in
the present case after the replacement $z\rightarrow \rho $ and $a=x^0$, $%
\theta $, $\varphi $, i.e., the axially-symmetric element is written in
spherical coordinates 
\begin{equation}
ds^2=e^{2u}\left( dx^0\right) ^2-e^{-2u}\left[ e^{2k}\left( d\rho ^2+\rho
^2d\theta ^2\right) +\rho ^2\sin ^2\theta d\varphi ^2\right] .  \label{hten}
\end{equation}
Expressions (66,67) for $S_{\;0\text{ }}^0$ and $S_{\;\varphi }^\varphi $
hold after the same change, while $S_{\;\theta }^\theta =0$. Once again we
argue that there is no mass surface layer, but only a charged one. As in the
previous case, for one sign of $q$ gravity becomes repulsive. The
interaction between massive bodies with repulsive (negative mass) and
attractive (positive mass) gravitation has been discussed in Refs \cite
{hone,htwo}. In our case repulsion is a natural property of the electric
field and does not break the energy conditions.

Let us deform the charged sphere into an oblate or prolate spheroid. Its
gravitational field is described best in the corresponding spheroidal
coordinates $x,y$ (to be distinguished from the cartesian coordinates in the
previous sections) 
\begin{equation}
r=\tau \left( x^2\pm 1\right) ^{1/2}\left( 1-y^2\right) ^{1/2},\quad z=\tau
xy,  \label{heleven}
\end{equation}
where $\tau $ is a parameter. In the prolate case one has 
\begin{equation}
x=\frac 1{2\tau }\left( l_{+}+l_{-}\right) \equiv \frac L\tau ,\quad y=\frac 
1{2\tau }\left( l_{+}-l_{-}\right) ,  \label{htwelve}
\end{equation}
\begin{equation}
l_{\pm }=\sqrt{\left( z\pm \tau \right) ^2+r^2}.  \label{hthirt}
\end{equation}
The master potential is taken to depend only on $x$ ($x=x_1$ is the surface
of the charged spheroid). The harmonic solution is given by the Legendre
function $Q_0\left( x\right) $ 
\begin{equation}
\psi =\frac q2\ln \frac{x-1}{x+1}  \label{hfourt}
\end{equation}
and generates the general solution of the Einstein-Maxwell equations for
such symmetry. From Eq. (44) one has $q=4\pi \sigma \left( x_1^2-1\right) $.
This potential coincides in form with the Weyl rod for the vacuum $\gamma $%
-metric \cite{twthree}. Eq. (34) gives 
\begin{equation}
k=\frac{Dq^2}8\ln \frac{x^2-1}{x^2-y^2},  \label{hfift}
\end{equation}
which depends on $y$ too, signalling non-inheritance, except when $B=2$. The
same conclusion follows in the oblate case, but there $\psi =-q\arctan 1/x$.
Repulsive gravity also arises in these cases.

It is well-known that the unique spherically symmetric electrovac solution
with mass $M$ and charge $\bar Q$ is the Reissner-Nordstr\"om solution \cite
{hthree,hfour} 
\begin{equation}
ds^2=\left( 1-\frac{2m}R+\frac{Q^2}{R^2}\right) c^2dT^2-\left( 1-\frac{2m}R+%
\frac{Q^2}{R^2}\right) ^{-1}dR^2-R^2\left( d\Theta ^2+\sin ^2\Theta d\Phi
^2\right) ,  \label{hsixt}
\end{equation}
\begin{equation}
\phi =\frac QR,\quad m=\frac{GM}{c^2},\quad Q^2=\frac{G\bar Q^2}{c^4}.
\label{hsevent}
\end{equation}
Hence, at least the solution given by Eq. (85) should be transformable into
the RN solution. Surprisingly, all three solutions described above are
formally equivalent to its three cases; undercharged ($Q^2<m^2$), critically
(extremely) charged ($Q^2=m^2$) and overcharged ($Q^2>m^2$). The proof uses
the Weyl form of the RN solution \cite{foseven,ninine}. The essential step
is to apply Kar's gauge, in which $\phi $ becomes harmonic instead of $\psi $%
. When $D=0$ we transform Eq. (85) by setting $R=\rho -q$, $\cos \Theta
=z/\rho $ and obtain Eqs. (93,94) with $m=-q$, $Q=q$. The sphere $\rho =\rho
_1$ transforms into the sphere $R=R_1=\rho _1-q$. When $D>0$ we take Eq.
(91) and fix $q$ by the condition $Dq^2/4=1$. Then we utilize Eqs. (38,39)
to find $\phi $ and $f$ 
\begin{equation}
\phi =-\frac 2{\sqrt{D}\left( \frac B{\sqrt{D}}+x\right) },\quad f=\frac{%
x^2-1}{\left( \frac B{\sqrt{D}}+x\right) ^2}.  \label{heightt}
\end{equation}
Making the identifications 
\begin{equation}
\tau =\left( m^2-Q^2\right) ^{1/2},\quad B=-\frac{2m}Q,  \label{hninet}
\end{equation}
we obtain (together with Eqs. (88-92)) the formulas for the undercharged
case from Ref. \cite{foseven}. The transformation 
\begin{equation}
R=\tau x+m,\quad \cos \Theta =y  \label{htwenty}
\end{equation}
maps this solution into the RN solution. The charged spheroid $x=x_1$ goes
into the charged sphere $R=R_1=\tau x_1+m$. It is seen from Eq. (96) that $%
B\rightarrow \infty $ when $Q\rightarrow 0$ and $B\phi $ remains finite. The
independent mass parameter $m$ again arises from the ''mass out of charge''
mechanism. A similar chain of arguments connects the oblate spheroid case to
the overcharged RN solution.

In conclusion, the electrically induced spherically symmetric gravitational
field is given by the critically charged Curzon solution, which is
equivalent to an extreme RN solution with $m=-Q$. There is a transformation
of RN solutions with $m\neq \pm Q$ into spheroidal metrics with particular $%
D $. The distance between the spheroid's foci $\tau $ is related to the
deviation of $m$ from its electromagnetic value. Thus part of the exterior
solution's mass is from electromagnetic origin. This has been long known for
interior charged perfect fluid solutions \cite{siseven}. The results are
analogous to the plane-symmetric ones. The usual RN solution with mass and
charge also exerts a repulsive force for certain values of its parameters 
\cite{hfive,hsix}. It appears, however, due to another mechanism. In the
charged Curzon solution the sign of the charge is decisive and the region
with repulsion occupies the whole exterior space. In the RN solution the
mass is always positive, but enters the metric with a negative sign, so that
a competition with the charge term is possible. The region with repulsion is
finite, $R<Q^2/m$.

\section{Magnetostatic examples}

We have pointed out that magnetic fields produce the same effects as
electric ones. One must replace $\phi $ by $\lambda $, $E_i$ by $H_i$, while 
$\psi $ becomes the magnetic scalar potential, determined by surface
currents. This is confirmed also by the definition of the magnetic field 
\cite{hfift} 
\begin{equation}
H_i=-\frac 12\sqrt{-g}\varepsilon _{ikl}F^{kl},  \label{hfoone}
\end{equation}
which gives, using Eq. (18) 
\begin{equation}
H_i=-\bar \lambda _i=-f\bar \psi _i.  \label{hfotwo}
\end{equation}
Eqs. (17,33) provide the connection between the scalar potential and the
only component ($\bar \chi =A_\varphi $) of the vector potential 
\begin{equation}
\bar \chi _z=r\bar \psi _r,\quad \bar \chi _r=-r\bar \psi _z.
\label{hfothree}
\end{equation}
It should be noted that in flat spacetime magnetostatics the formula $\vec H%
=curl\vec a$ involves the physical component in curvilinear coordinates $%
a_{\left( \varphi \right) }=A_\varphi /r$. In magnetogravity one has to find
the scalar potential anyway, in order to obtain $f$.

In a general medium the energy is $T_{00}\sim \mu H^2$, where $\mu $ denotes
now the magnetic constant and consequently $T_{\mu \nu }\rightarrow \mu T_{\mu \nu }$%
. Hence, $\psi $ picks also the multiplier $\sqrt{\mu }$. Eq. (99) remains
unchanged. Arguments, analogous to those for electric fields, lead to the
conclusion that $B=2$ for a pure magnetic effect upon gravity. Formula (45)
for the acceleration becomes with high degree of precision 
\begin{equation}
g_i=-\sqrt{G\mu }H_i.  \label{hfofour}
\end{equation}
It is similar to the electric case, Eqs. (49,71). The gravitational force is
very different from the Lorentz force, acting upon charged particles 
\begin{equation}
\vec F_L=e\vec E+\frac{e\mu }c\left( \vec v\times \vec H\right) .
\label{hfofive}
\end{equation}
In it the electric force does not involve $\varepsilon $, while the magnetic
does involve $\mu $, but acts only on moving charges. There are terms,
depending on the velocity, also in $g_i$. However, we are interested in the
gravitational effect upon macroscopic bodies for which $v\ll c$, hence, we
have neglected them and study only acceleration at rest.

The first magnetovac Weyl solution was given by Papapetrou \cite{twelve}.
Later Bonnor presented a number of examples \cite{five,twone}, including the
gravitational field of a current loop. He found a very small effect, based
on mass-energy considerations.

The magnetic field of a current loop is well-known and includes elliptic
integrals. On the $z$-axis and at the centre the field simplifies 
\begin{equation}
H_z\left( z,0\right) =0.2\pi I\frac{r_1^2}{\left( r_1^2+z^2\right) ^{3/2}}%
,\quad H_0\equiv H_z\left( 0,0\right) =\frac{0.2\pi I}{r_1},  \label{hfosix}
\end{equation}
where $r_1$ is the loop's radius in $cm$, $H$ is measured in Gauss and the
current $I$ is measured in $Amps$. Setting $\mu =1$ we get from Eq. (103) 
\begin{equation}
g_0\equiv \left| g_z\left( 0,0\right) \right| =1.62\times 10^{-4}\frac I{r_1}%
.  \label{hfoseven}
\end{equation}
Earth's acceleration is reached when $H_e=3.8\times 10^6G=380T$ or $%
I_e/r_1=6.05\times 10^6A/cm$. For a laboratory set-up let us take $r_1=100cm$%
. If a lightning with cross-section of radius $r_0=10cm$ circles around the
loop, one may take $I=10^5A$, the current density being $J=3.18\times
10^2A/cm^2$ and $g_0=0.162cm/s^2$. One can reach $g_e$ with a current of $%
6.05\times 10^8A$.

It is advantageous to make the loop thicker, turning it into a finite
solenoid. Let the inner radius be $r_1$, the outer radius be $r_2$ and the
height be $l$. Then \cite{hsixt,hsevent} 
\begin{equation}
H_0=F\left( \alpha ,\beta \right) r_1J,\quad F\left( \alpha ,\beta \right)
=0.4\pi \beta \ln \frac{\alpha +\sqrt{\alpha ^2+\beta ^2}}{1+\sqrt{1+\beta ^2%
}},  \label{hfoeight}
\end{equation}
where $\alpha =r_2/r_1$, $\beta =l/r_1$ and $J$ is the current density. As
an example, let us take $r_1=100cm$, $r_2=2r_1$, $l=2r_1$. Then $F\approx 1$
and $J=H_0/100$. Now $g_e$ is reached when $J_e=3.8\times 10^4A/cm^2$.

One can increase the acceleration by creating magnetic fields in a
ferromagnetic medium. Iron has $\mu _{\max }=5000$ ($\sqrt{\mu _{\max }}%
=70.7 $). There are alloys, like supermalloy, which have $\mu _{\max
}=8\times 10^5 $, $\sqrt{\mu _{\max }}=894.4$. Their saturation field is
comparatively low, $H_s=8\times 10^3G$ and the maximum is obtained roughly
for one third of this value. The effective field in Eq. (101) will be $%
H_{eff}=\sqrt{\mu _{\max }}H_{\max }\approx 238T$, which is of the order of $%
H_e$. Making a disc from this material with a current flowing in a strip
around the rim, one gets the magnetic analogue of the moving capacitor
example from Sec.V.A.

There are two issues which need clarification. The metric depends directly
on the scalar potential, but $\psi $ is multivalued in magnetostatics. If
there is a current-carrying surface, the jump takes place there and probably
the metric can be made continuous in this region of non-vanishing $J$. In
the case of a current loop the jump can be arranged to take place on any
surface, based on the loop. Symmetry considerations require to take the disk 
$z=0$, $\rho <r_1$ as such a surface. In spherical coordinates \cite{twone} 
\begin{equation}
\psi =0.2\pi J\left[ \pm 1-\frac \rho {r_1}P_1+\frac{\rho ^3}{2r_1^3}%
P_3-...+\left( -1\right) ^{n+1}\frac{1.3...\left( 2n-1\right) }{2.4...2n}%
\left( \frac \rho {r_1}\right) ^{2n+1}P_{2n+1}+...\right] ,  \label{hfonine}
\end{equation}
where the sign coincides with $signz$ and $P_n\left( \cos \theta \right) $
are the Legendre polynomials. Now, let us remember that we have been unable
to determine the sign of $B$ and for definiteness worked with positive $B$.
In fact, Eq. (35) should read 
\begin{equation}
f=\left( 1\pm \psi \right) ^{-2}.  \label{hfifty}
\end{equation}
Using different signs for $z$ positive or negative, and taking into account
that $P_{2n+1}\left( \cos \frac \pi 2\right) =0$, makes $f$ continuous at $%
z=0$. The derivatives of $\psi $ are single-valued, hence, $k$ is also
continuous. The acceleration will have a jump and change of sign at the disk
within the loop, like it has on both sides of a charged plane (disk). We
took the absolute value of $z$ in Eq.(60), which plays the same role. This
jump has been attributed to some mass surface layer \cite{twone}, which in
our view is not correct.

The second issue concerns the fact that the magnetic field far away from the
loop has a dipole character and the acceleration does not contain a monopole
term. As we have argued, this is not a tragedy, because electromagnetic
fields can create artificial gravity in a confined region of space.
Intuitively, this is more favourable energetically. Of course, the material
of the loop always has some mass, which induces a monopole term, negligible
with respect to the strong effect from root gravity.

\section{Conclusions}

The results of this paper can be summarised as follows. Some of them are not
new and the corresponding references are cited in these cases.

1) The gravitational acceleration at rest in Weyl-Majumdar-Papapetrou fields
has a root gravity term, proportional to $c^2\sqrt{\kappa }=\sqrt{G}$, which
is $10^{23}$ times bigger than the usual perturbative coefficient $c^2\kappa 
$. It is linear in the EM-fields, while the perturbative term is quadratic.
Sizeable gravitational force exists (see Eq. (49)) although the metric is
very close to the flat one (Maxwellian limit). Its explicit form determines
up to a sign the important constant $B$. For its typical value $B=2$ the
Earth's acceleration is obtained in electric fields of order $10^9V/cm$ and
in magnetic fields of about $380T$. One can change the direction of $g_i$ by
changing the direction of $E_i$ or $H_i$ and confine $g_i$ to a finite
volume by confining the EM-fields.

2) The energy-momentum tensor (in particular, its energy component $T_{00}$)
induces a change in the Ricci tensor $\sim \kappa $ according to the
Einstein equations (3). This leads to changes in the metric and its
acceleration, which can be $\sim \sqrt{\kappa }$ and may contain no monopole
term. Then the formula $E=mc^2$ does not hold. Creating artificial gravity
that is localized in space and has no long-distance mass terms seems
energetically more favourable.

3) In axially-symmetric systems Weyl fields provide regular exterior and
interior solutions to any distribution of charges or currents on a closed
surface. They are determined by a master-potential $\psi $ satisfying the
Laplace equation. When the metric depends on just one coordinate (three
commuting Killing vectors are present) the Weyl solution becomes the most
general one. In truly axisymmetric cases it presumably determines the pure
electromagnetic effect on gravity, while the solutions of the Ernst equation
include hidden mass sources. The constant $B$ (taken positive for
definiteness) divides the Weyl solutions into three classes, according to $%
B=2,$ $B>2$, $B<2$. \cite{ninet}. Among them $B=2$ is privileged, being the
simplest (conformastatic spacetimes). When $B\neq 2$, parasitic masses
appear. In some cases the metric does not inherit the symmetry of the
EM-source.

4) The gravitational force at rest induced by electric and magnetic fields
is the same \cite{fift,sixt,sevent} unlike the Lorentz force, acting upon
charged particles. The surface sources determine the master potential and
not $A_\mu $ \cite{five,ninet}.

5) There is a ''mass out of charge'' mechanism, which allows to obtain
solutions with mass from the Weyl solutions. It clearly indicates the part
of mass which is of electromagnetic origin. Eq. (12) still holds and root
gravity term remains, but $B$ is affected by the mass. Such solutions can
incorporate the mass of the charged surface, which is always present in
practice.

6) In the general static case a harmonic master potential for WMP fields
appears only when $B=2$ \cite{eleven,twelve}. Point or line sources have
singularities, hence, one should use closed surface sources (shells).
Another alternative is to use volumes of charged dust or perfect fluid,
where the functional dependence $f\left( \phi \right) $ appears naturally as
an equilibrium condition. For charged dust a harmonic potential can always
be introduced, but the equilibrium is unstable \cite{fifour,fieight}. It is
more realistic to charge conductive surfaces by applying potentials or pass
currents through coils wound around them, relying on the motion of free
electrons in metals. One can also create powerful fields by separating the
positive and negative ions.

7) In order to construct global solutions around the charged shells, the
fulfilment of the junction conditions is required. They have some subtleties
in the Weyl case. Eq. (12) interrelates the Einstein and the Maxwell
equations (27) and a jump in $T_{\mu \nu }$ results due to the jump in $\psi 
$, caused by the charges or the currents. A mass surface layer is
not necessary. The acceleration is caused by the EM-field and not by often
unrealistic fluid sources, trying to duplicate this effect.

8) The pure electric plane-symmetric effect on the metric is described by
the McVittie solution, which is a Weyl solution of class $B=2$. When $B\neq
2 $ the Weyl metric is not plane-symmetric (non-inheritance of the source's
symmetry). Solutions with mass and charge also contain a root gravity term.

9) A usual freely hanging capacitor may be used to test root gravity.
Constant acceleration is induced in its dielectric, Eq. (71). Taking maximum
potential difference, minimum distance between the plates and a material
with very big dielectric constant (ferroelectric) one can obtain $g_z=5.2g_e$%
. This force is trying to move the capacitor in the $z$ direction. It is
smaller than the force of electric attraction between the plates.

10) The pure electric spherically-symmetric effect on the metric is
described by the charged Curzon solution, Eq. (85), which is related by a
simple transformation to the critically (extremely) charged RN solution.
This is a Weyl solution with $B=2$. When $B\neq 2$ non-inheritance of the
source's symmetry results. The same is true for prolate and oblate
spheroidal solutions. They are singular at the axis, so one can use them as
exteriors to a charged spheroidal shell with Minkowski interior. Their
gravity becomes repulsive in the whole outside region for one sign of the
charge. Typically, the repulsive $g=0.5cm/s^2$. There is a formal coordinate
transformation between RN fields with mass and charge and Weyl fields
outside spheroidal charged shells.

11) The magnetic field of a current loop induces a gravitational force
according to Eq. (104). The Earth's acceleration is reached when the current 
$I=6\times 10^8A$. A thick loop (fat solenoid) requires for this goal
current density of $3.8\times 10^4A/cm^2$. A ferromagnetic disc with a
current strip around it is the magnetic analogue of the moving capacitor.

\section{Discussion}

Root gravity has been overlooked in the past. One reason is the wide-spread
use of relativistic units - already Weyl worked in them. The only papers on
the subject (except the present one) written in non-relativistic units are
by Ehlers \cite{thnine,forty}, but are extremely rarely cited.

Another reason is that the simpler solutions with plane, spherical,
spheroidal or cylindrical symmetry were studied directly and were almost
never treated as subcases of the axisymmetric solutions. This cut their ties
to Weyl solutions from the very beginning. The RN solution was shown to
belong to the Weyl class only in 1972 \cite{foseven}.

A third reason is the underestimation of exact solutions in favour of
approximation schemes and numerical techniques. Root gravity escapes
undetected by these methods. In fact, it sets a new scale $\sqrt{\kappa }$
or $\sqrt{G}$ and perturbations should be done around the exact Weyl
solutions. WMP solutions have never been in the main trend of development of
general relativity. The fact that the unique charged black hole is a WMP
solution pushed their investigation towards multiple-charged-black holes 
\cite{eleven,twelve,thirt,twenty,foseven,foeight,fonine}, particle
trajectories in these fields \cite{htwnine} or WMP-based wormholes \cite
{hthirty}. A similar astrophysical issue are charged dust clouds (Bonnor
stars) and their non-singular interpolation to black holes \cite
{fifour,fieight,hthone,hthtwo}. Axisymmetric solutions have been studied
mainly with the generation techniques for the Ernst equation. Papers on WMP
solutions often appeared in local, old or unavailable journals.

A fourth reason concerns the motivation of the first researchers. Weyl
himself was much more interested in his conformal theory of gravity, trying
to unify gravitation and electromagnetism. He never returned back to the
Weyl fields. McVittie presented his solution as an example to test one of
the unified theories of Einstein. The latter concentrated his energy on the
geometrical unification of the two fundamental long-range interactions.

In this paper we share another view, closer to the Rainich ''already unified
theory''. We are interested in the pure electromagnetic effects upon
gravity. It is not necessary to use his formalism, which includes products
of Ricci tensors and is similar in complexity to the Gauss-Bonnet terms in
the string effective action. The point is that we know how to create strong
EM-fields. Our approach is not only to explain but to construct and
parallels the efforts to build wormholes and warp drives. We use, however,
fields which are common and satisfy all energy conditions and not exotic
matter or the Casimir effect. Second, we try to induce gravitational force
without altering very much the metric. This is possible thanks to the 20
orders of magnitude collected in $c^2/2$ in Eq. (7). We have demonstrated
that except the Newtonian limit, general relativity possesses also a
Maxwellian limit. Root gravity is one way to do this, but there are others
too.

We have given here several laboratory set-ups, where root gravity can be
detected and general relativity tested once more, this time in its highly
non-linear regime. The most promising seems to be the well-known capacitor
and its magnetic analogue. The effect has not been discovered yet, because
capacitors are used in low-voltage circuits and they are first firmly fixed
and charged afterwards. Small root gravity effects should be present also in
today's solenoids with strong magnetic fields. The acceleration is masked by
that of the Earth, being a fraction of it, and is hardly measurable because
the working space is very small.

\end{document}